# Critical behavior for impact fragmentation of spherical solid bodies sensitive to strain rate

N.N. Myagkov[1]

*Institute of Applied Mechanics, Russian Academy of Sciences, Leningradsky Prospect 7, Moscow 125040, Russian Federation*

## Abstract

We consider the impact fragmentation of two spherical solid bodies sensitive to strain rate in a three-dimensional (3D) setting. We use both dimensional analysis and numerical simulations by smoothed-particle hydrodynamics (SPH) method to shed light on this problem. The key point of the work is the assumption of complete self-similarity of the problem under consideration with respect to the effective strain rate parameter $\dot{\varepsilon}_{eff}$, which is verified by numerical simulations. As a result we consider the two cases corresponding to the high-velocity $\dot{\varepsilon}_{eff} \gg 1$ and low-velocity $\dot{\varepsilon}_{eff} \ll 1$ loading. The size of the system may be characterized by the total number of the SPH particles $N_{tot}$ approximating each sphere. It is shown that for finite system the critical velocity of fragmentation at high-velocity loading exceeds that at low-velocity loading, i.e. $V_{c\infty} > V_{c0}$. With an unlimited increase in the system size these velocities become the same. It is shown that $V_{c\infty}$ and $V_{c0}$ depend on the system size in a quadratic manner, i.e. $V_c^2 - V_c^2(\infty) \propto N_{tot}^{-1/3\nu}$ where $\nu$ is a correlation length exponent .

## 1. INTRODUCTION

Phenomena of dynamic fragmentation are widespread in nature, they occur, for example, in the collision of heavy nuclei in atomic physics [1, 2], the collision of macroscopic bodies [3-9] (including the impact of a striker on a thin or massive target).

A possible critical behavior during fragmentation was considered both in the problem of nuclear collisions at moderate energies [1, 2], and in the problems of fragmentation of mechanical systems [5, 6, 10-15]. An approach was used based on the similarity of the observed fragmentation parameters, for example, the mass distribution of fragments, and the results of known theories of critical phenomena, for example, percolation. The hypothesis of the presence of a damage-fragmentation phase transition was verified for mechanical systems of different types both by experimental studies [11, 14] and by computer ones [5, 6, 15].

---

[1] E-mail: n.myagkov@iam.rus.ru; Phone: +7 495 9461765 (office); orcid.org/0000-0001-9293-7834

Despite the large amount of experimental and theoretical works, many questions of the critical behavior remain open. One of these is the critical behavior for impact fragmentation of solids sensitive to strain rate. This paper attempts to shed light on this problem using both dimensional analysis and numerical simulations by smoothed particle hydrodynamics (SPH) method.

Molecular dynamics (MD) [12, 16, 18], discrete element models (DEM) [5, 6, 17], and SPH method [10, 13, 15] are used to simulate fragmentation numerically. The choice of the particle-based MD and DEM simulations to analyze the impact fragmentation is motivated by lack of reliable models and complexity of describing effects from the continuum point of view, namely plasticity, fracture and phase transitions. Meanwhile these effects are automatically obtained by integration of the motion equations in case of MD method with the aid of an interaction potential only. In the framework of DEM, the solid is represented as an assembly of cohesive elements making possible a realistic treatment of materials' microstructure and of the loading conditions. The use of the SPH method makes it possible to conduct the simulations based on the complete system of equations of deformed solid mechanics (DSM).

An important feature of the SPH method is that a proper SPH lattice should be as regular as possible and not contain large discrepancies in order to perform most accurate simulation [19, 20]. Therefore, in the SPH calculation, the computational domain is discretized, as a rule, by a cubic lattice with SPH particles located at the lattice nodes.

The problem of impact fragmentation of two identical aluminum spheres of diameter $D$ moving towards each other along a straight line connecting their centers is considered in this paper. The module of the initial velocity of each sphere $V$ relative to the center of mass of the spheres is interpreted below as the impact velocity. Due to the symmetry of the problem, we consider the fragmentation of one sphere by placing a plane of symmetry in the center of mass of the spheres. Therefore all data, which are shown in the paper, are data of one sphere.

We use dimensional analysis [21, 22] for studying the scaling properties of the sphere fragmentation. Also the sphere fragmentation is simulated by the 3D numerical solution by the SPH method [20]. The Mie - Gruneisen equation of state and the Johnson - Cook (JC) model for the yield stress are used to describe the material behavior of the spheres. The use of the JC model allows us to take into account the system response on the strain rate.

## 2. DIMENSIONAL ANALYSIS AND SCALING

## 2.1. Rigid-plastic model of material

Consider the simplest model of material, which is characterized by a rigid-plastic response to external influences. It is described by two parameters: density $\rho_0$ and yield strength $Y_0$. Besides the problem includes the diameter of sphere $D$, the initial impact velocity $V$ and the period $a$ of the cubic lattice of the SPH particles, by which the sphere is discretized. Notice that the parameters $\rho_0$, $Y_0$ and $V$ are not dimensionally independent so it is impossible to obtain the parameter of length out of them. It means that the solution of the problem (e.g., the average fragment mass $M_{avr}$) depends on parameters $K = \dfrac{\rho_0 V^2}{Y_0}$ and $\dfrac{a}{D}$. Taking into account that $\dfrac{a}{D} \approx (N_{tot})^{-1/3}$ we obtain a functional dependence of the form

$$M_{avr} = m_{tot} \cdot F_1\left(K, N_{tot}^{-1}\right), \qquad (1)$$

where $m_{tot} = \dfrac{1}{6}\pi\rho_0 D^3$ is the initial mass of the sphere and $N_{tot}$ is the total number of the SPH particles in the sphere.

We see (1) that the natural parameter characterizing the size of the system is $N_{tot}$. Moreover, the change of $N_{tot}$ can occur not only due to the change in the diameter of the sphere $D$, but also due to the change in the lattice period $a$. However, we consider that passage to the limit $N_{tot} \to \infty$ means $D \to \infty$, but not $a \to 0$.

Calculations of the average fragment mass $M_{avr}$ are usually performed to investigate the damage-fragmentation transition when the impact velocity changes [5, 6, 10, 13, 15]. It is defined as the average

$$M_{avr} = \langle M_2^j / M_1^j \rangle \qquad (2)$$

where $M_1^j$ and $M_2^j$ are the first and second moments of fragment mass distribution $M_k^j(V) = \sum_m m^k n^j(m,V) - m_{max,j}^k$ in the $j$th simulation, $n^j(m,V)$ is the number of fragments with mass $m$ in the $j$th simulation at the impact velocity $V$ and the sum runs over all fragments. Note that the contribution of the largest fragment of mass $m_{max,j}$ is subtracted from $M_k^j(V)$. Angle brackets $\langle \ldots \rangle$ in (2) denote averaging over an ensemble of simulations for each value of $V$. In this paper at least 12 simulations were performed for each value of $V$. The simulations differed from each other by perturbations introduced into the initial conditions through angular displacement around the rotation axis of the SPH-discretized sphere.

The critical impact velocity $V_c$ is determined as follows: for $V < V_c$ the damage occurs, and for $V > V_c$ the fragmentation takes place. As is known [5, 6, 10, 13, 15], $V_c$ (or introduced above parameter $K_c = \dfrac{\rho_0 V_c^2}{Y_0}$) is defined as the position of the maximum of the dependence of $M_{avr}$ on $V$ (or $K$). Then from (1) we have

$$K_c = F_2(N_{tot}^{-1}) \tag{3}$$

From (3) it can be supposed that at $N_{tot} \to \infty$

$$K_c(N_{tot}) = K_c(\infty) + \dfrac{A}{N_{tot}^\lambda}. \tag{4}$$

where $\lambda$ is a critical exponent [23], $K_c(\infty) = F_2(0)$ and $A$ = const. The equation (4) estimates $K_c(N_{tot})$ in a finite sphere with $N_{tot}$ particles versus $K_c(\infty)$ for infinite sphere with $N_{tot} \to \infty$. Note that the limit transition $N_{tot} \to \infty$ in (4) can only correspond to $D \to \infty$ but not $a \to 0$. If we put exponent $\lambda = 1/(3\nu)$, the equation (4) takes the form of equation $p_{av} - p_c \propto L^{-1/\nu}$ from [24], which is used in percolation theory to estimate the percolation threshold $p_{av}$ in a finite system of size $L$ versus the percolation threshold $p_c$ for infinite system where the exponent $\nu$ makes sense of a correlation length exponent. The equation of the form (4) was before used in [6] to analyze fragmentation in a similar problem. However, in contrast to equation (4), the authors of [6] write this equation as $V_c - V_c(\infty) \propto D^{-1/\nu}$, i.e. they postulate a linear dependence on $V_c$.

Let the problem also depend on the fragment mass $m$. Then the average cumulative mass distribution $N_{avr}$ depends on parameters $m$, $\rho_0$, $Y_0$, $V$, $D$ and $a$. We choose three dimensionally independent parameters $\rho_0$, $V$, and $D$, then we obtain

$$N_{avr} = F_3\left(K, \dfrac{m}{m_{tot}}, N_{tot}^{-1}\right) \tag{5}$$

From (3) and (5) it follows that the average cumulative mass distribution at the critical point $K = K_c$ depends only on $m/m_{tot}$ and $N_{tot}$.

Obviously, the described scaling allows the problem of finding $M_{avr}$ and $N_{avr}$ to be solved depending on $D$ only for one lattice period $a$. The dependences $M_{avr}$ and $N_{avr}$ on $D$ for other values of $a$ can be obtained by simple recalculation from the solution found.

**2.2. Model of material sensitive to strain rate**

A more realistic model of material, in addition to the parameters of density $\rho_0$ and yield strength $Y_0$ must to take into account the elastic modulus $E$ (or the adiabatic bulk modulus) and shear modulus $G$, specific heat $C_V$, the initial temperature $T_0$ and melting temperature $T_m$, the dynamic ultimate tensile strength $\sigma_p$, and a strain-rate parameter $\dot{\varepsilon}_0$, which is the parameter of constitutive equation and, generally speaking, is determined from experiment. As before we take into account the impact velocity $V$, lattice period $a$ and the sphere diameter $D$. We choose four parameters with independent dimensions $Y_0$, $V$, $D$ and $T_0$. Then, similarly to (1), we can write

$$M_{avr} = m_{tot} \cdot Q_1\left(K, N_{tot}^{-1}, \dot{\varepsilon}_{eff}, \Pi\right) \tag{6}$$

where $\dot{\varepsilon}_{eff} = \dfrac{V}{\dot{\varepsilon}_0 D}$ is the effective strain rate during the impact, $\Pi = \left\{ \dfrac{C_V T_0 \rho_0}{Y_0}, \dfrac{T_m}{T_0}, \dfrac{\sigma_p}{Y_0}, \dfrac{E}{Y_0}, \dfrac{G}{Y_0} \right\}$ is the set of dimensionless parameters independent of $V$ and $D$, and $K = \dfrac{\rho_0 V^2}{Y_0}$. In (6), it was taken into account that $\dfrac{C_V T_0}{V^2} = \dfrac{C_V T_0 \rho_0}{Y_0} \cdot K^{-1}$. Similarly to (3) and (5) we have

$$K_c = Q_2(N_{tot}^{-1}, \dot{\varepsilon}_{eff}, \Pi) \tag{7}$$

$$N_{avr} = Q_3\left(K, \dfrac{m}{m_{tot}}, N_{tot}^{-1}, \dot{\varepsilon}_{eff}, \Pi\right) \tag{8}$$

We now consider the parameter $\dot{\varepsilon}_{eff}$ in $Q_i$ ($i = 1,2,3$) (6)-(8). On the physical level of rigor, some dimensionless parameter is considered significant, i.e. determining a phenomenon in question, if it is not too large and not too small. Otherwise, it is assume that the influence of this parameter can be neglected. This reasoning is indeed valid if the finite limits of the functions $Q_i$ in (6-8) as the parameter $\dot{\varepsilon}_{eff}$ tending to zero or infinity exist with the other similarity parameters remaining constant. In general, in practice, for sufficiently small or sufficiently large $\dot{\varepsilon}_{eff}$ the functions $Q_i$ can be replaced with the required accuracy by the functions including the arguments without $\dot{\varepsilon}_{eff}$. In other words, we assume the complete self-similarity of the problem with respect to the parameter $\dot{\varepsilon}_{eff}$ [22]. Thus, from (6) - (8) we have

$$\begin{aligned} (M_{avr})_\infty &= m_{tot} \cdot Q_1\left(K, N_{tot}^{-1}, \infty, \Pi\right) \\ (M_{avr})_0 &= m_{tot} \cdot Q_1\left(K, N_{tot}^{-1}, 0, \Pi\right) \end{aligned} \tag{9}$$

$$\begin{aligned} K_{c\infty} &= Q_2(N_{tot}^{-1}, \infty, \Pi) \\ K_{c0} &= Q_2(N_{tot}^{-1}, 0, \Pi) \end{aligned} \tag{10}$$

$$(N_{avr})_\infty = Q_3\left(K, \frac{m}{m_{tot}}, N_{tot}^{-1}, \infty, \Pi\right)$$
$$(N_{avr})_0 = Q_3\left(K, \frac{m}{m_{tot}}, N_{tot}^{-1}, 0, \Pi\right) \tag{11}$$

In (9)-(11) we denoted the limiting cases corresponding to the high-velocity $\dot{\varepsilon}_{eff} \gg 1$ and low-velocity $\dot{\varepsilon}_{eff} \ll 1$ loading by the lower indices $\infty$ and 0, respectively. Here, one can see an analogy with the mechanics of nonequilibrium processes [25], where the terms "frozen" and "equilibrium" process are used for such limiting cases.

It can be seen that $(M_{avr})_\infty$, $(N_{avr})_\infty$, $(M_{avr})_0$ and $(N_{avr})_0$ at the fixed set of dimensionless parameters $\Pi$ have the same scale invariance property as relations (1) and (5).

As will be seen from the simulations for finite $N_{tot}$, the critical velocity of fragmentation at high-velocity loading exceeds that at low-velocity loading, i.e.

$$K_{c\infty} > K_{c0} \quad \text{or} \quad V_{c\infty} > V_{c0} \tag{12}$$

From (10), similarly to (4), we have

$$K_{c\infty}(N_{tot}) = K_{c\infty}(\infty) + \frac{A_\infty}{N_{tot}^{1/3\nu}},$$
$$K_{c0}(N_{tot}) = K_{c0}(\infty) + \frac{A_0}{N_{tot}^{1/3\nu}}, \tag{13}$$

where $A_\infty$ and $A_0$ do not depend on $N_{tot}$; $K_{c0}(\infty) = \frac{\rho_0 V_{c0}^2(\infty)}{Y_0}$, and $K_{c\infty}(\infty)$ is defined similarly. Let's define $V_{c0}(\infty)$ and $V_{c\infty}(\infty)$ as the critical velocity of fragmentation for a sphere with unlimited growth of $N_{tot}$ and $D$.

It can be seen from the definition $\dot{\varepsilon}_{eff} = \frac{V}{\dot{\varepsilon}_0 D}$ that no matter how small we take $\dot{\varepsilon}_0$, initially large $\dot{\varepsilon}_{eff}$ becomes small with the growth of $D$ at $N_{tot} \to \infty$. Thus, it makes sense to consider large effective strain rates $\dot{\varepsilon}_{eff} \gg 1$ as an intermediate asymptotics for spheres with $N_{tot} \ll (V/a\dot{\varepsilon}_0)^3$ (or the diameter $D \ll V/\dot{\varepsilon}_0$) for a sufficiently small $\dot{\varepsilon}_0$. In this case, $V_{c\infty}(\infty)$ should be defined as the critical fragmentation velocity at $\overline{N}_{tot}$ (or $\overline{D}$) such that

$$(N_{tot})_{init} \ll \overline{N}_{tot} \ll (V/a\dot{\varepsilon}_0)^3 \quad \text{or} \quad D_{init} \ll \overline{D} \ll V/\dot{\varepsilon}_0 \tag{14}$$

where $(N_{tot})_{init}$ (or $D_{init}$) is the largest value of $N_{tot}$ (or $D$) at which the fragmentation calculations were performed in this work.

## 3. NUMERICAL SIMULATIONS

### 3.1. Numerical simulation method and material model

The problem of impact fragmentation of two identical aluminum spheres of diameter $D$ moving towards each other along a straight line connecting their centers with the same initial velocity module $V$ is considered in this paper. Due to the symmetry of the problem, we consider the fragmentation of one sphere and all data, which are shown in the paper, are data for one sphere.

The 3D numerical simulations were based on the DSM equations and performed by the SPH method implemented in LS-DYNA version 971 program package [20]. We perfomed the calculations for spheres with diameters $D$ = 6.35, 8.10, 9.6, 11.19 and 12.27 mm with the number of SPH particles $N_{tot}$ = 17269, 35825, 59757, 94533 and 124800 ( $N_{tot}$ corresponds to one sphere).. The simulation method and the algorithm for finding fragments are described in more detail by us in [10, 13, 15].

The material models are similar to those that we used in Ref. [13]. Mie-Gruneisen equation of state and Johnson-Cook model [26]

$$Y = \left(Y_0 + B(\bar{\varepsilon}^p)^n\right) \cdot (1 + C \ln \dot{\varepsilon}^*) \cdot \left(1 - (T^*)^m\right) \qquad (15)$$

for the yield strength were taken as the constitutive equations. In (15) it is denoted: $\bar{\varepsilon}^p$ is the equivalent plastic strain; $\dot{\varepsilon}^* = \dot{\bar{\varepsilon}}^p / \dot{\varepsilon}_0$ is the dimensionless equivalent plastic strain rate where $\dot{\varepsilon}_0$ is the parameter which is specified in the simulation parameters; $T^* = (T - T_0)/(T_m - T_0)$.

Table 1 shows the aluminum alloy data that we used in the calculations, where $E$ is the elastic modulus, $G$ is the shear modulus, $\rho_0$ is the initial density of the material, $k$ is the factor in the shock adiabat $D_{sw}=c_0+k*U$, $\Gamma$ is the Gruneisen coefficient, which was assumed to be constant, and $\sigma_p$ is the ultimate tensile strength. The maximum principal stress spall model was used as the fracture model in the numerical simulations. This simple fracture model that is independent of the strain rate allowed us to keep the computational time within reasonable limits. For example, the use of the Johnson-Cook fracture model leads to a significant increase in the computation time, and the solution of the problem under consideration becomes excessively time-consuming.

The parameters $Y_0$, $B$, $C$, $n$, $m$ and $T_m$ in (15) are material parameters. The values of material parameters in (15) used in numerical simulations were taken from Ref. [13]. The simulations were carried out for two values of the parameter $\dot{\varepsilon}_0$: $\dot{\varepsilon}_0 = 1$ s$^{-1}$ and $10^6$ s$^{-1}$. The characteristic impact velocity is $V \approx 0.6$ km/s. The sphere diameters $D$ for which the simulations were carried out are given above. Thus, these two values $\dot{\varepsilon}_0$ correspond to the effective strain rates $\dot{\varepsilon}_{eff} = \dfrac{V}{\dot{\varepsilon}_0 D}$ approximately equal to $\dot{\varepsilon}_{eff} \approx 5 \cdot 10^4$ and $5 \cdot 10^{-2}$.

Tab. 1. Material parameters

| Material | $\rho_0$, g/cm$^3$ | $E$, GPa | $G$, GPa | $\sigma_p$, GPa | $k$ | $\Gamma$ | $C_V$ kJ/(kg·°K) | $T_m$, °K |
|---|---|---|---|---|---|---|---|---|
| Aluminum alloy | 2.71 | 72.8 | 27.3 | 1.15 | 1.34 | 2.0 | 0.875 | 875 |

### 3.2 Results of numerical simulations.

In this section, through numerical simulation, we verify the assumptions and conclusions made in section 2.

1. The complete self-similarity with respect to the effective strain rate $\dot{\varepsilon}_{eff}$.

With the help of a numerical solution it is impossible to simulate the passage to the limit $\dot{\varepsilon}_{eff} \to \infty$ or $\dot{\varepsilon}_{eff} \to 0$. However, it can be shown that for small and large $\dot{\varepsilon}_{eff}$ the values of $(M_{avr})_\infty$, $(N_{avr})_\infty$, $(M_{avr})_0$ and $(N_{avr})_0$ in (9) - (11) do not tend to zero or infinity and take different finite values.

The calculation results for $D = 9.6$ mm ($N_{tot} = 59757$) are shown in Fig. 1a,b for the average fragment mass, and the average cumulative distribution of the fragments at the critical impact velocities $V_{c\infty}$ and $V_{c0}$. It can be seen that $(M_{avr})_\infty$ and $(M_{avr})_0$ as well as $(N_{avr})_\infty$ and $(N_{avr})_0$ differ from each other, taking finite values. The positions of the sharp maximums in Fig. 1 (a) are the critical values of the impact velocity $V_{c\infty}$ and $V_{c0}$. They separate the damage phase from the fragmentation phase. It is seen that $(N_{avr})_\infty$ and $(N_{avr})_0$ in Fig. 1 (b) are close over the entire range of mass variation. For spheres of other diameters, the calculations give the similar result.

Noteworthy is the presence of humps in curves $(M_{avr})_\infty$ and $(M_{avr})_0$ (Fig. 1a), which are located in the supercritical region ($V > V_c$). The humps in Fig. 1a and almost horizontal distributions in the region of average masses in Fig. 1b, which are plotted at critical impact velocities, are related. From the distribution in Fig. 1b it can be seen that there are a large number of small fragments and, in addition, there are the largest (dominant) fragments having a dimensionless mass $m/m_{tot}$ of the order O(1). At $V > V_c$, these dominant fragments are fractured;

as a result, the hump appears in the dependences of the average fragment mass on the impact velocity. This is confirmed by the distributions $(N_{avr})_\infty$ and $(N_{avr})_0$ in Fig. 1c, plotted at impact velocities of 0.6 and 0.58 km/s corresponding to the hump centers for $\dot{\varepsilon}_{eff} \gg 1$ and $\dot{\varepsilon}_{eff} \ll 1$ (shown by arrows in Fig. 1a). In Fig. 1c, we already observe the almost power-law distributions in the region of average masses.

The values of $V_{c0}$ and $V_{c\infty}$ found are shown in Tab. 2 as a function of the sphere diameter $D$ and the number of SPH particles $N_{tot}$ used in the simulation of the sphere. It can be seen that inequality (8), $V_{c\infty} > V_{c0}$, holds for all $N_{tot}$. Thus, the simulations confirm the hypothesis of the complete self-similarity with respect to the effective strain rate $\dot{\varepsilon}_{eff}$ at least on the physical level of rigor.

Tab. 2. Results of determining the critical impact velocities $V_{c0}$ and $V_{c\infty}$.

| $N_{tot}$ | 17269 | 35825 | 59757 | 94533 | 124800 |
|---|---|---|---|---|---|
| $D$, мм | 6.35 | 8.10 | 9.60 | 11.19 | 12.27 |
| $V_{c\infty}$ (km/s) | 0.646±0.001 | 0.578±0.001 | 0.536±0.001 | 0.515±0.001 | 0.504±0.001 |
| $V_{c0}$ (km/s) | 0.628±0.001 | 0.563±0.001 | 0.523±0.001 | 0.503±0.001 | 0.494±0.001 |

It can be seen that the difference between $V_{c0}$ and $V_{c\infty}$ is not great. This is due to the smallness of the coefficient $C$ in front of the term that takes into account the strain rate in the Johnson-Cook model (15). The coefficient $C$ varies from ≤0.01 to 0.017 depending on the type of aluminum alloy (see, for example, [26-28]). In the simulations, we took $C$ = 0.01 in accordance with the data [26].

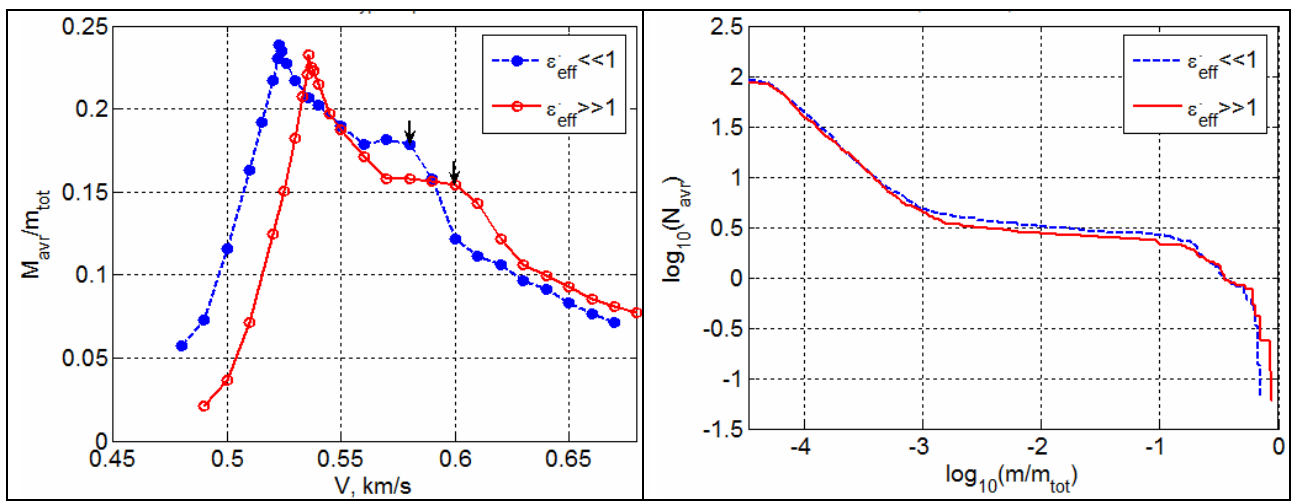

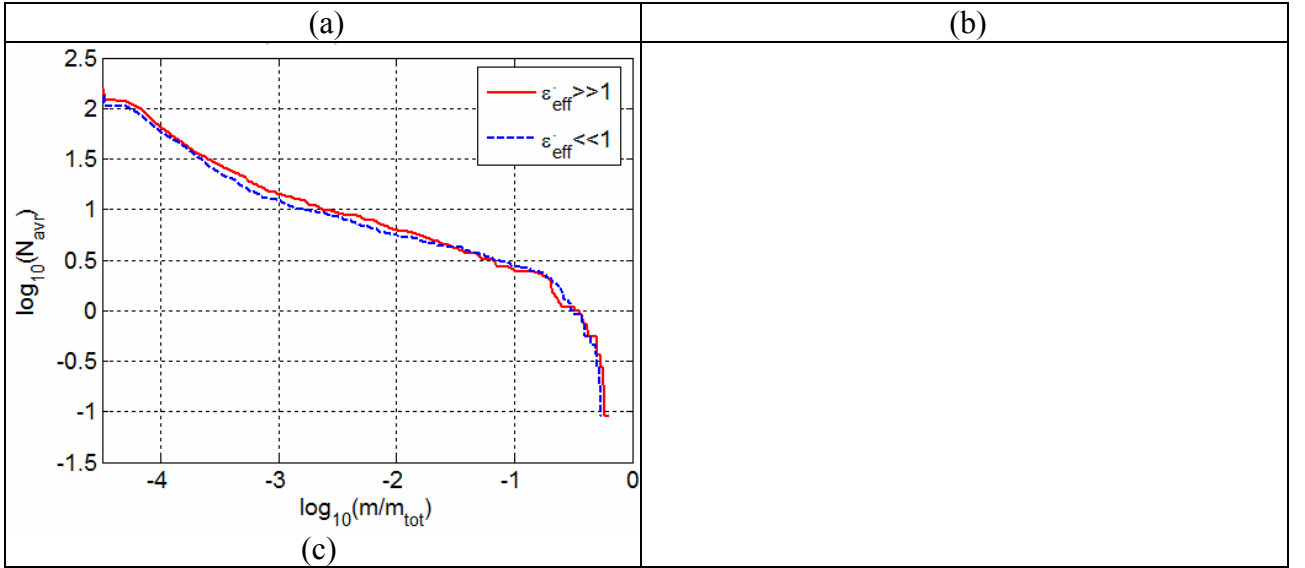

Fig. 1. The fragmentation at the high-velocity $\dot{\varepsilon}_{eff} \approx 5 \cdot 10^4$ and low-velocity $\dot{\varepsilon}_{eff} \approx 5 \cdot 10^{-2}$ loading. In figure: (a) Normalized average fragment mass $(M_{avr})_\infty / m_{tot}$ and $(M_{avr})_0 / m_{tot}$ as a function of impact velocity $V$; (b) Average cumulative mass distribution of fragments $(N_{avr})_\infty$ and $(N_{avr})_0$ at critical impact velocities $V_{c\infty} = 0.536$ km/s and $V_{c0} = 0.523$ km/s; (c) $(N_{avr})_\infty$ and $(N_{avr})_0$ at impact velocities of 0.6 and 0.58 km / s (shown by arrows in Fig. 1a). Diameter of the sphere $D = 9.6$ mm, the number of SPH particles $N_{tot} = 59757$.

2. We verify the scaling by changing the lattice period $a$ and the sphere diameter $D$, determining $(M_{avr})_\infty$ and $(N_{avr})_\infty$, for a fixed value of the number of SPH particles $N_{tot}$ and various values of $a$ and $D$. The simulation results for $N_{tot} = 59757$ are presented in Fig. 2. It can be seen that the simulations for spheres of different diameters, but with the same number of SPH particles ($N_{tot} = 59757$), are in good agreement. For other values of $N_{tot}$ the similar picture is observed.

Therefore, the effects accompanying fragmentation can, generally speaking, be associated not with the sphere diameter $D$, but with the number of SPH particles $N_{tot}$ in this sphere. However, unlimited growth of the sphere at the limit transition $N_{tot} \to \infty$ should be associated with $D \to \infty$ for fixed value of the lattice period $a$.

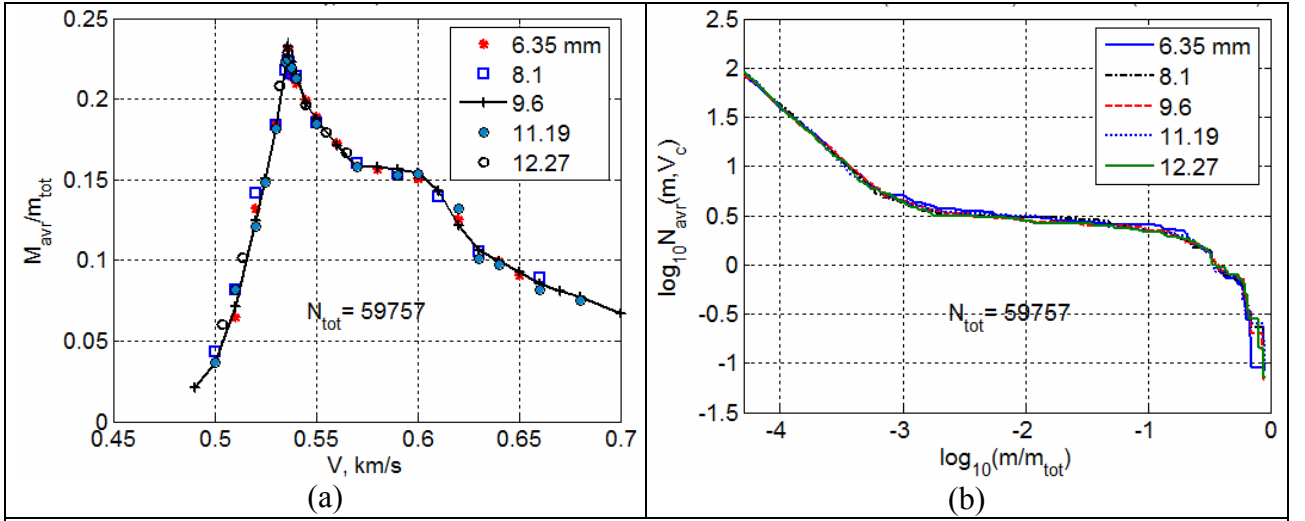

Fig. 2. The fragmentation at the high-velocity ($\dot{\varepsilon}_{eff} \approx 5 \cdot 10^4$) loading. Normalized average masses of fragments $(M_{avr})_\infty / m_{tot}$ as a function of impact velocity $V$ (a) and average cumulative mass distributions of fragments $(N_{avr})_\infty$ at critical impact velocities $V_{c\infty} = 0.536$ km/s (b) for spheres of different diameters $D$ (shown in the figure), but with the same value $N_{tot} = 59757$.

3. We treat the formulas (13) rewriting them in the form

$$V_{c\infty}^2 = V_{c\infty}^2(\infty) + \frac{A'_\infty}{N_{tot}^{1/3\nu}}$$

$$V_{c0}^2 = V_{c0}^2(\infty) + \frac{A'_0}{N_{tot}^{1/3\nu}}, \qquad (16)$$

where $V_{c0}$ and $V_{c\infty}$ as functions of $N_{tot}$ are given in Table 2. In (16) the critical velocities $V_{c0}(\infty)$ and $V_{c\infty}(\infty)$ are the fitting parameters. Their values are determined by the least squares method for each of the two equations in (16) (see Fig. 3). The best agreement of $V_{c0}$ and $V_{c\infty}$ with the calculated data, as expected, is obtained with the same value of $V_c(\infty)$ equal to 0.446 km/s. While the correlation length exponent $\nu = 0.459 \pm 0.015$. Note that for 3D percolation in the site problem $\nu \approx 1.0$ [24].

Let us evaluate how the value of $V_{c\infty}(\infty)$ obtained above correspond to the constraint (14). From Table 2 it can be seen that $(N_{tot})_{init} = 124800$. Let us take some value $\overline{N}_{tot}$ a thousand times less than the upper limit in (14): $\overline{N}_{tot} = 0.001 \cdot (V/a\dot{\varepsilon}_0)^3 \sim 1.5 \cdot 10^{16}$ ($V \approx 0.5$ km/s, $a \approx 0.2$ mm, $\dot{\varepsilon}_0 = 1$ s$^{-1}$). It is seen that condition (14) $(N_{tot})_{init} \ll \overline{N}_{tot} \ll (V/a\dot{\varepsilon}_0)^3$ is satisfied. Now from (16) for this $\overline{N}_{tot}$ taking into account that $A'_\infty \approx 300$ we find: $V_{c\infty}^2(\overline{N}_{tot}) - V_c^2(\infty) = \frac{A'_\infty}{\overline{N}_{tot}^{1/3\nu}} \sim 10^{-10}$, i.e. $V_{c\infty}(\overline{N}_{tot})$ really close to the limit value $V_c(\infty)$.

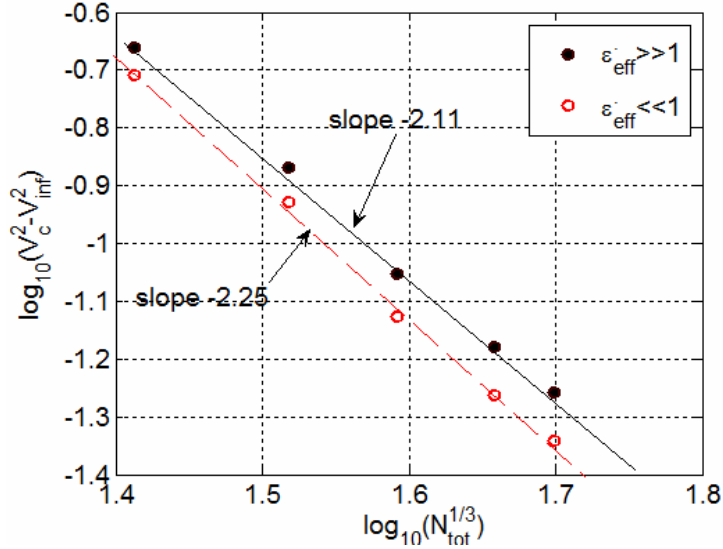

Fig. 3. The difference between the squares of the critical impact velocities of the finite and infinite systems $V_c^2(N_{tot}) - V_c^2(\infty)$ as a function of the size of the system $N_{tot}$. In the figure: $V_{inf}$ denotes $V_c(\infty)$, ● and ○ are the calculated points and – - and —— are the approximating straight lines.

## 4. Conclusions

We consider the impact fragmentation of two spherical solid bodies sensitive to strain rate in a three-dimensional (3D) setting. These are two identical aluminum spheres that move towards each other along the straight line connecting their centers with the same velocity module $V$. Due to the symmetry of the problem, we consider the fragmentation of one sphere.

Numerical simulations of the fragmentation are perfomed by the SPH method based on the complete system of the DSM equations. Dimensional analysis for extracting the scale properties of fragmentation is applied by us to the discretized sphere which has a new parameter with the dimension of length, namely the lattice period.

Mie-Gruneisen equation of state and Johnson-Cook (JC) model [26] for the yield strength were taken as the constitutive equations. The use of the JC model allows us to take into account the the mechanical response on the strain rate.

At first we considered the simplest model of material, which is characterized by a rigid-plastic response to external influences. The rigid-plastic material is described by a very limited set of material parameters: density ρ and yield strength $Y$. Nevertheless already here, by means of elementary analysis, we find that the solutions of the problem (e.g., the average fragment mass (2), the average cumulative mass distribution (5), etc.) depend on the square of the impact velocity and the total number of the SPH particles $N_{tot}$ that approximate the sphere. Note that dependence (4), which estimates the critical velocity of fragmentation for the finite system, generally speaking follows from the scaling for $M_{avr}$ (1):

$$M_{avr} = m_{tot} \cdot N_{tot}^{\gamma/3\nu} \widehat{F}_1\left([K - K_c(\infty)]N_{tot}^{1/3\nu}\right) \tag{17}$$

where $K = \dfrac{\rho_0 V^2}{Y_0}$ and $K_c = \dfrac{\rho_0 V_c^2}{Y_0}$, $\nu$ and $\gamma$ are the critical exponents, and $\widehat{F}_1(\cdot)$ is the some function. The connection between equations of the form (4) and (17) was discussed in the book [29]. Scaling of the form (17) was before used in [6] to analyze the impact fragmentation in the similar problem. However, in [6], in contrast to (17), the linear dependence of the argument on the impact velocity $V$ was postulated.

The use of the JC model leads to the fact that among the dimensionless parameters of the problem a dimensionless effective strain rate $\dot{\varepsilon}_{eff} = \dfrac{V}{\varepsilon_0 D}$ appears. To simplify the problem, we assume the complete self-similarity of the problem with respect to the parameter $\dot{\varepsilon}_{eff}$ (about the complete self-similarity see, for example, [22]), which assumes the existence of the problem solution in the two limiting cases corresponding to the high-velocity $\dot{\varepsilon}_{eff} \gg 1$ and low-velocity $\dot{\varepsilon}_{eff} \ll 1$ loading (they are denoted the lower indices ∞ and 0, respectively). The fact that these limiting solutions exist is confirmed by numerical simulation. Of course, with the help of a numerical solution it is impossible to simulate the passage to the limit $\dot{\varepsilon}_{eff} \to \infty$ or $\dot{\varepsilon}_{eff} \to 0$. However, we show that for sufficiently small and large $\dot{\varepsilon}_{eff}$ the values of $(M_{avr})_\infty$, $(N_{avr})_\infty$, $(M_{avr})_0$ and $(N_{avr})_0$ in (9) - (11) do not tend to zero or infinity and take different finite values. In addition, instead of one dependence (4), which estimates the critical fragmentation velocity $V_c$ for the finite system, two similar dependences (13) appear for the high-frequency $V_{c\infty}$ and low-frequency $V_{c0}$ critical velocities.

Briefly, the main results of the work are as follows:
- The key assumption of complete self-similarity with respect to the effective strain rate parameter $\dot{\varepsilon}_{eff}$ is confirmed by numerical simulations at least at the physical level of rigor.
- The scaling by changing the lattice period $a$ and the sphere diameter $D$ is demonstrated. Therefore, the effects accompanying fragmentation should, generally speaking, be associated not with the sphere diameter $D$, but with the number of SPH particles $N_{tot}$ in this sphere ($\dfrac{a}{D} \approx (N_{tot})^{-1/3}$).

- Numerical simulations show that for finite $N_{tot}$ the critical velocity of fragmentation at high-velocity loading exceeds that at low-velocity loading, i.e. $V_{c\infty} > V_{c0}$. With the growth of $N_{tot}$ and $D$, $V_{c\infty}$ and $V_{c0}$ take the same limit value $V_c(\infty)$ = 0.446 km/s. While the correlation length exponent ν = 0.459±0.015.
- It is shown that $V_{c\infty}$ and $V_{c0}$ depend on the system size in a quadratic manner, i.e. $V_{c\infty}^2 - V_c^2(\infty) \propto N_{tot}^{-1/3\nu}$ and $V_{c0}^2 - V_c^2(\infty) \propto N_{tot}^{-1/3\nu}$ where ν is a correlation length exponent.

**Acknowledgments**